\documentclass{iaus}
\usepackage{graphicx}

\title
[Galaxy Interactions, Star Formation History, and Bulgeless Galaxies]
{Galaxy Interactions, Star Formation History, and Bulgeless Galaxies}

\author[Shardha Jogee]   
{Shardha Jogee$^1$}
\affiliation{$^1$
Department of Astronomy, University of Texas at Austin, \\
1 University Station C1400, Austin, TX 78712-0259 \\
email: {\tt sj@astro.as.utexas.edu}
}

\pubyear{2008}
\volume{254}  
\pagerange{1--6}
\jname{The Galaxy Disk in Cosmological Context}
\editors{A.C. Editor, B.D. Editor \& C.E. Editor, eds.}
\begin{document}

\maketitle

\begin{abstract}
Hierarchical $\Lambda$CDM models provide a successful paradigm for the growth of dark matter 
on large scales, but they face important challenges in predicting how the baryonic components of galaxies evolve. I present constraints on two aspects of this evolution: (1) The interaction history of galaxies over the last 7 Gyr and the impact of interactions on their star formation properties, based on Jogee et al. (2008a,b);  (2)~Constraints on  the origin of bulges in  hierarchical  models and the challenge posed in accounting for galaxies with low bulge-to-total ratios, based on Weinzirl, Jogee, Khochar, Burkert, \& Kormendy (2008, hereafter WJKBK08).

\end{abstract}

\firstsection 
\section{Galaxy Interactions and their Star Formation over the Last 7 Gyr}

The merger history of galaxies impacts the mass assembly (e.g., 
Dickinson et al. 2003), star  formation history, AGN activity (e.g., Springel. 
et al. 2005b)  and  structural evolution of galaxies.  
The merger rate/fraction at $z>1$ remains highly uncertain, owing 
to relatively modest volumes and bandpass shifting effects, but with a 
general trend towards higher merger fractions at higher redshifts 
Even the merger rate at $z<1$~has  proved hard to robustly measure for a variety 
of reasons, ranging from small samples in early studies, to different
methods on large samples in later studies.

In Jogee et al. (2008a,b), we have  performed a  complementary and
comprehensive observational estimate of the frequency of interacting 
galaxies over  $z \sim$~0.24--0.80 (lookback times of 3--7 Gyr), and 
the  impact of interactions on the star formation (SF)  of  galaxies over
this interval. Our study is based on  $HST$ ACS, COMBO-17, and Spitzer 24~$\mu$m data 
from the GEMS survey. 
We use a large sample of $\sim$~3600 ($M \ge$~$1 \times  10^{9}$ $M_{\odot}$)  galaxies 
and  $\sim$~790 high mass  ($M \ge$~$2.5 \times 10^{10}$ $M_{\odot}$) galaxies for 
robust number statistics.
Two independent methods  are used to identify strongly interacting galaxies:  a tailored 
visual classification system complemented with spectrophotometric redshifts and stellar
masses, as well as the CAS merger criterion 
($A >$~0.35 and $A > S$; Conselice 2003), based on CAS asymmetry $A$ and clumpiness 
$S$ parameters.
This allows one of the most extensive comparisons to date between CAS-based and visual 
classification results.
We set up this visual classification system  so as  to target interacting systems  whose  
morphology and other properties  suggest they are  a recent merger of mass ratio $M1/M2>$~1/10.  
While many earlier studies focused only on major mergers, we try  to constrain the
frequency of minor mergers as well, since they dominate 
the merger rates in $\Lambda$CDM models. Some of our  results are outlined below.

\vspace{0.5mm}
\bf (1) \rm
Among $\sim$~790 high  mass  galaxies, {\it the fraction of  visually-classified 
interacting systems over lookback times of 3--7 Gyr ranges from  9\% $\pm$~5\% at 
$z \sim$~0.24--0.34, to 8\% $\pm$~2\%  at $z \sim$~0.60--0.80, as averaged over 
every Gyr bin.(Fig.~1a).}  
These systems  appear to be in merging or post-merger phases, and are candidates 
for a recent merger of mass ratio $M1/M2>$~1/10.
Similar  results on the interaction fraction are 
reported  by Lotz et al. (2008). 
The  lower limit on the major  ($M1/M2 >$ 1/4) merger  fraction
ranges from  1.1\%  to 3.5\%  over $z\sim$~0.24--0.80.
The corresponding lower limit on the minor  (1/10 $\le M1/M2 <$ 1/4)  
merger fraction ranges  from  3.6\% to 7.5\%.  This is the first, albeit approximate, 
empirical estimate of the frequency of  minor mergers over the last 7 Gyr.

\vspace{0.5mm}
\bf (2) \rm
For an assumed value of $\sim$~0.5 Gyr for the visibility timescale, it follows 
that 
{\it  each massive ($M \ge$~$2.5 \times  10^{10}$ $M_{\odot}$) galaxy has  
undergone $\sim$~0.7 mergers of mass ratio~$>$~1/10 over  the redshift interval 
$z\sim$~0.24--0.80. } Of these, we estimate that 1/4 are major mergers, 
2/3 are minor mergers, and the rest are  ambiguous cases of  major or minor mergers.
The corresponding merger rate $R$ is a few $\times 10^{-4}$ galaxies 
Gyr$^{-1}$ Mpc$^{-3}$.
Among $\sim$~2840 blue cloud galaxies of mass $M \ge$~$1.0 \times 10^{9}$ $M_{\odot}$, 
similar results hold.

\vspace{0.5mm}
\bf (3) \rm
We compare our empirical merger rate $R$ for high mass 
($M \ge$~$2.5 \times  10^{10}$ $M_{\odot}$) galaxies 
to predictions from different  $\Lambda$CDM-based simulations  of galaxy evolution,
including  the halo occupation distribution (HOD) models of Hopkins et al.  (2007); 
semi-analytic models (SAMs) of Somerville et al.  (2008), 
Bower et al. (2006), and Khochfar \&  Silk (2006); and smoothed 
particle hydrodynamics (SPH) cosmological simulations from Maller et al.  (2006). 
To our knowledge, such extensive comparisons have not been attempted to date, and 
are long overdue.
{\it We find  qualitative  agreement  between the observations and models, with
the (major+minor) merger  rate from different models bracketing  
the observed rate, and showing a factor of five dispersion} (Fig.~1b).
One can now anticipate that in the near future, improvements in both the observational 
estimates and model predictions will start to rule out certain merger scenarios and refine
our understanding of the merger history of galaxies.

\vspace{0.5mm}
\bf (4) \rm
The idea that galaxy interactions generally enhance the star formation rate (SFR) of galaxies  is well established
from  observations (e.g., Joseph \& Wright 1985; 
Kennicutt et al. 1987)  and simulations 
(e.g., Hernquist 1989; Mihos \& Hernquist 1994, 1996;  Springel, Di Matteo \& Hernquist 2005b). 
However, simulations cannot uniquely predict the  factor by which  interaction enhance
the SF activity of galaxies over the last 7 Gyr, since both the SFR and properties 
of the remnants in simulations are highly sensitive to the  stellar feedback model,  
the bulge-to-disk ($B/D$) ratio, the gas mass fractions, and orbital geometry   
(e.g., Cox et al 2006; di Matteo et al. 2007).   
Thus, empirical constraints are needed.
Among $\sim$~3600 intermediate mass ($M \ge$~$1.0 \times 10^{9}$ $M_{\odot}$) galaxies, 
we find that 
{\it 
the average SFR  of visibly interacting galaxies is  
only modestly enhanced compared to non-interacting galaxies over $z \sim$~0.24--0.80} 
(Fig.~1c).
This result is found for SFRs based on UV, UV+IR, and UV+stacked-IR data.
This modest enhancement is consistent with the results  of di 
Matteo et al. (2007) based on numerical simulations of several hundred  
galaxy collisions.

\vspace{0.5mm}
\bf (5) \rm
The SF properties of interacting and non-interacting galaxies since $z< 1$ are of great 
astrophysical interest, given that the  
cosmic SFR density is claimed to decline by a factor of 4 to 10 since $z\sim$~1 (e.g., 
Lilly et al. 1996; Ellis et al 1996;  Hopkins 
2004; P{\'e}rez-Gonz{\'a}lez  et al.\ 2005;  Le Floc'h et  al.\ 2005).
We therefore set quantitative limits on the  contribution of 
obviously interacting  systems to the UV-based and  UV+IR-based SFR density  
over $z\sim$~0.24--0.80.
Among  $\sim$~3600 intermediate mass ($M \ge$~$1.0 \times 10^{9}$ $M_{\odot}$)  galaxies, 
we find  that 
{\it visibly interacting systems only account for a small fraction ($<$ 30\%) of 
the cosmic SFR density over lookback times of $\sim$~3--7 Gyr ($z\sim$~0.24--0.80; 
Fig.~(1d))}. 
Our result is consistent with that of Wolf et al. (2005) 
over a smaller lookback time interval of  $\sim$~6.2--6.8 Gyr. In effect, our result 
suggests that {\it the behavior  of the cosmic SFR density over the  last 7 Gyr is  
predominantly shaped by non-interacting galaxies, rather than strongly interacting 
galaxies}. 
This suggests that the observed decline in the  cosmic  SFR density 
since $z \sim$~0.80  is largely the result of a  shutdown 
in the SF of non-interacting galaxies.

\section
{The origin of bulges and the problem of bulgeless galaxies}

In $\Lambda$CDM  models of galaxy evolution, there are in  principle 
three main mechanisms to build bulges of spiral galaxies: major mergers,  
minor mergers,  and  secular processes  (see WJKBK08 for details).
The major merger of two spiral galaxies destroys the disk component and leaves
behind a classical bulge, around which a stellar disk  forms when  
hot gas in the halo subsequently cools, settles into a disk, and forms stars. 
Minor mergers can also grow bulges in several  ways.  
A  tidally induced bar and/or  direct tidal torques from  the companion 
can drive gas into the inner kpc 
(e.g.,  Quinn et al. 1993;  Hernquist \& Mihos 1995;  Jogee 2006 and references therein), 
where subsequent SF  forms a compact  high $v/\sigma$  stellar component, or disky pseudobulge.
In addition, the stellar core of the satellite can sink to the central region via dynamical friction.
Finally, bulges can also have a secular origin: here, a stellar bar or globally oval 
structure in a  {\it non-interacting} galaxy drives  gas inflow into the inner kpc , where 
subsequent SF forms a disky pseudobulge (e.g., Kormendy 1993; Jogee 1999; Kormendy \& Kennicutt 
2004;  Jogee, Scoville, \& Kenney 2005).

These different  mechanisms to form bulges have been  postulated for a long time.
However, what is still missing is {\rm a quantitative assessment of the relative importance of different bulge 
formation  pathways} in high and low mass spirals.
For instance, although bulges are an integral part of massive present-day spiral 
galaxies, we still cannot answer the following basic question: do most bulges 
in massive spirals  form via  major mergers,  minor mergers, or secular processes? 

Another thorny issue is the prevalence of bulgeless galaxies.
There is rising evidence  that   bulgeless galaxies are quite 
common in the local Universe  (e.g., B{\"o}ker et al.   2002;
Kautsch et al.  2006;  BJM08a; Kormendy \& Fisher 2008).
Yet, in $\Lambda$CDM models of galaxy evolution, most galaxies that had a past 
major merger at a time when their mass was a fairly large fraction of their
present-day mass,  are expected to have a significant bulge.
So far, no quantitative comparisons have been done between observations  and model 
predictions to assess  how serious is the challenge  posed by 
bulgeless galaxies.

In  WJKBK08, we attempt one of the first  quantitative comparisons  of the properties  
of bulges in a fairly complete sample of high mass 
($M_\star \geq 1.0 \times 10^{10} M_\odot$) spirals 
to predictions from $\Lambda$CDM-based simulations of galaxy evolution.
We derive the  bulge-to-total mass ratio   ($B/T$)  and bulge 
S\'ersic index $n$  by performing 
$2D$ bulge-disk-bar decomposition on $H$-band images of 146 bright, high 
mass, moderately inclined spirals. 

\vspace{0.5mm}
\bf (1) \rm
Interestingly, we find that as many as $\sim$~56\% of 
high  mass spirals have low $n \leq 2$ bulges: such bulges exist  in barred and 
unbarred galaxies across all Hubble  types  (Fig.~2a).
Furthermore a striking  $\sim$~66\% of  high  mass spirals have  $B/T \le$~0.2 
(Figs.~3a and 3b).

\vspace{0.5mm}
\bf (2) \rm
We compare the observed  distribution of bulge $B/T$ in high mass spirals 
to predictions  from  $\Lambda$CDM-based  semi-analytical models. 
In the models, a bulge with  $B/T \le$~0.2 can exist in a galaxy with a past major 
merger, only if the last major  merger occurred at $z>2$ (lookback $>$~10 Gyr).  
The predicted fraction of high mass spirals with  a past major merger and 
a bulge with a present-day  $B/T \le 0.2$ is  {\it a factor of over 
fifteen smaller}  than the observed fraction ($\sim$~66\%) of high mass spirals
with $B/T \le 0.2$  (Fig.~2b). 
The comparisons {\it rule out major mergers as the main formation  pathway
for bulges in high mass spirals}. Contrary to common perception, 
{\it bulges built  via  major mergers seriously fail to account for the bulges 
present in $\sim$~66\% of  high  mass spirals}.

\vspace{0.5mm}
\bf (3) \rm
In the models, the majority of low $B/T \le$~0.2  bulges  exist in systems that 
have experienced {\it only minor mergers, and no major mergers} (Fig.~2b).  These bulges can 
be built  via minor mergers and secular processes. 
So far, we explored one realization of the model focusing on bulges built via satellite stars  
in minor mergers and find good agreement with the observations. Future models will 
explore more realistic minor merger scenarios and secular processes.

\clearpage
\vspace{-7.0 mm}    
\hspace{-0.5 cm}
\includegraphics[width=2.4 in]{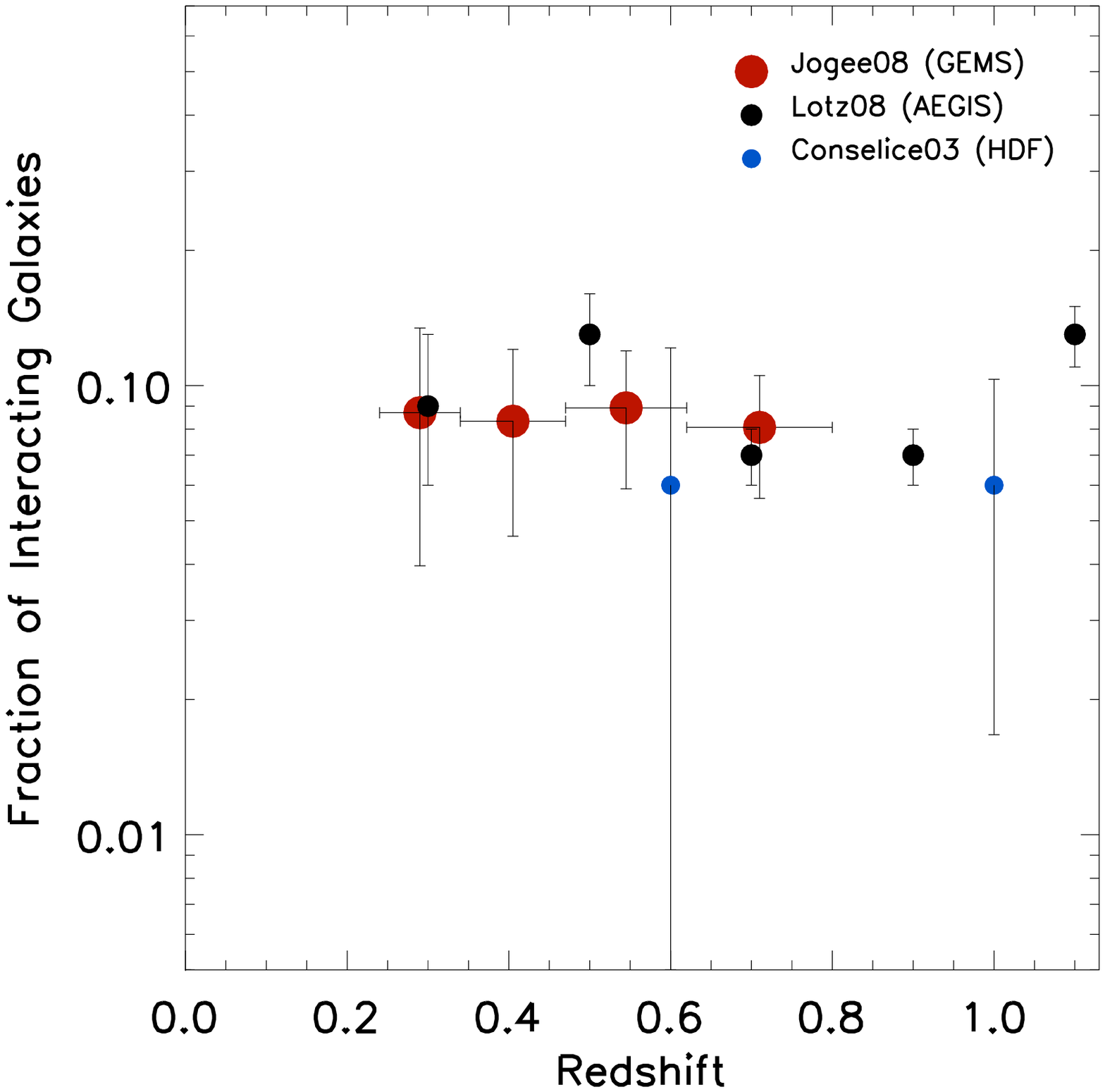}
\vspace {0.5 mm}
\hspace{3.0 mm}
\includegraphics[width=2.65in]{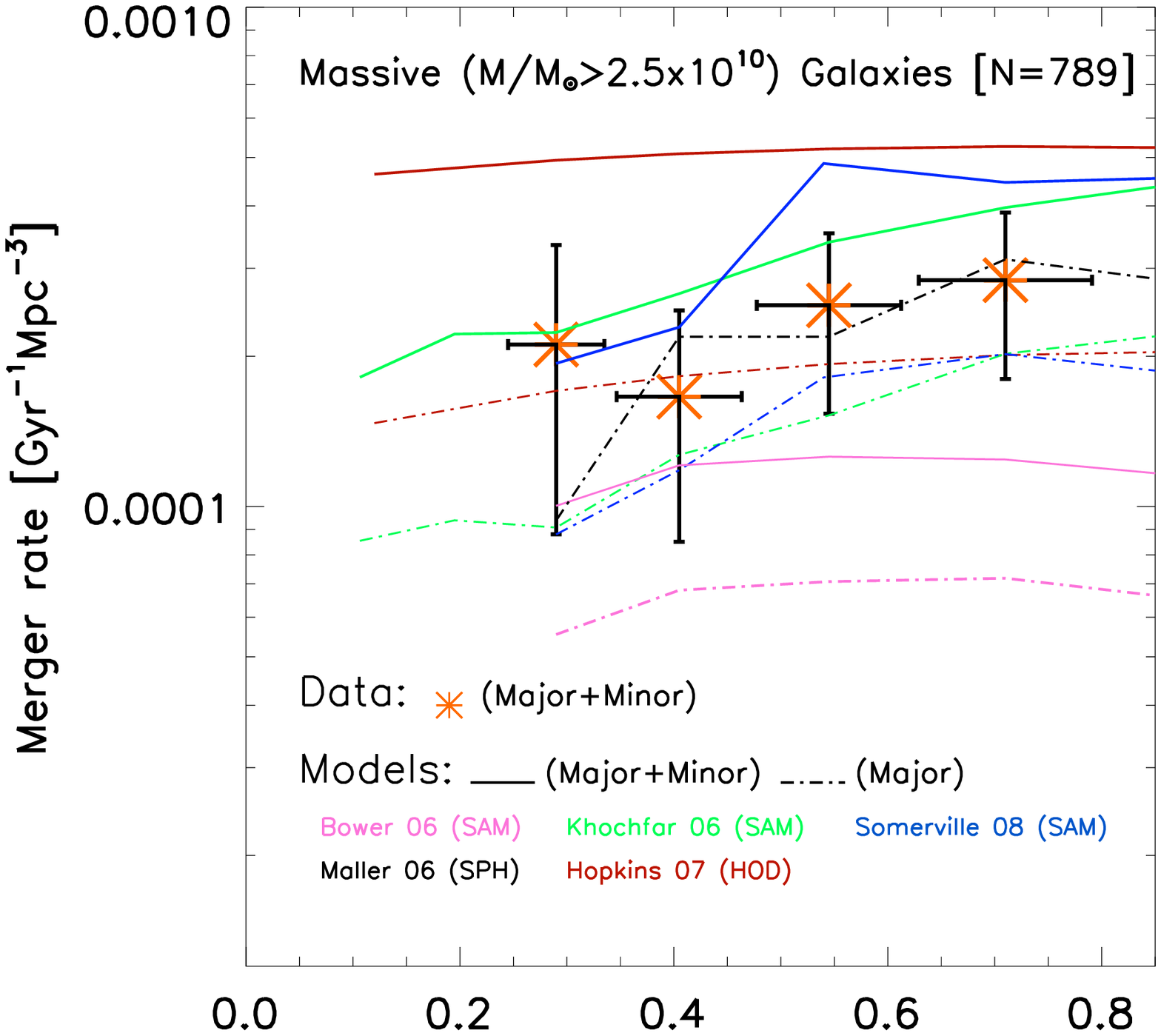}
\vskip 0.1 cm   
\vspace{0.0 mm}   
\hspace{-1.0 cm}
\includegraphics[width=2.7 in]{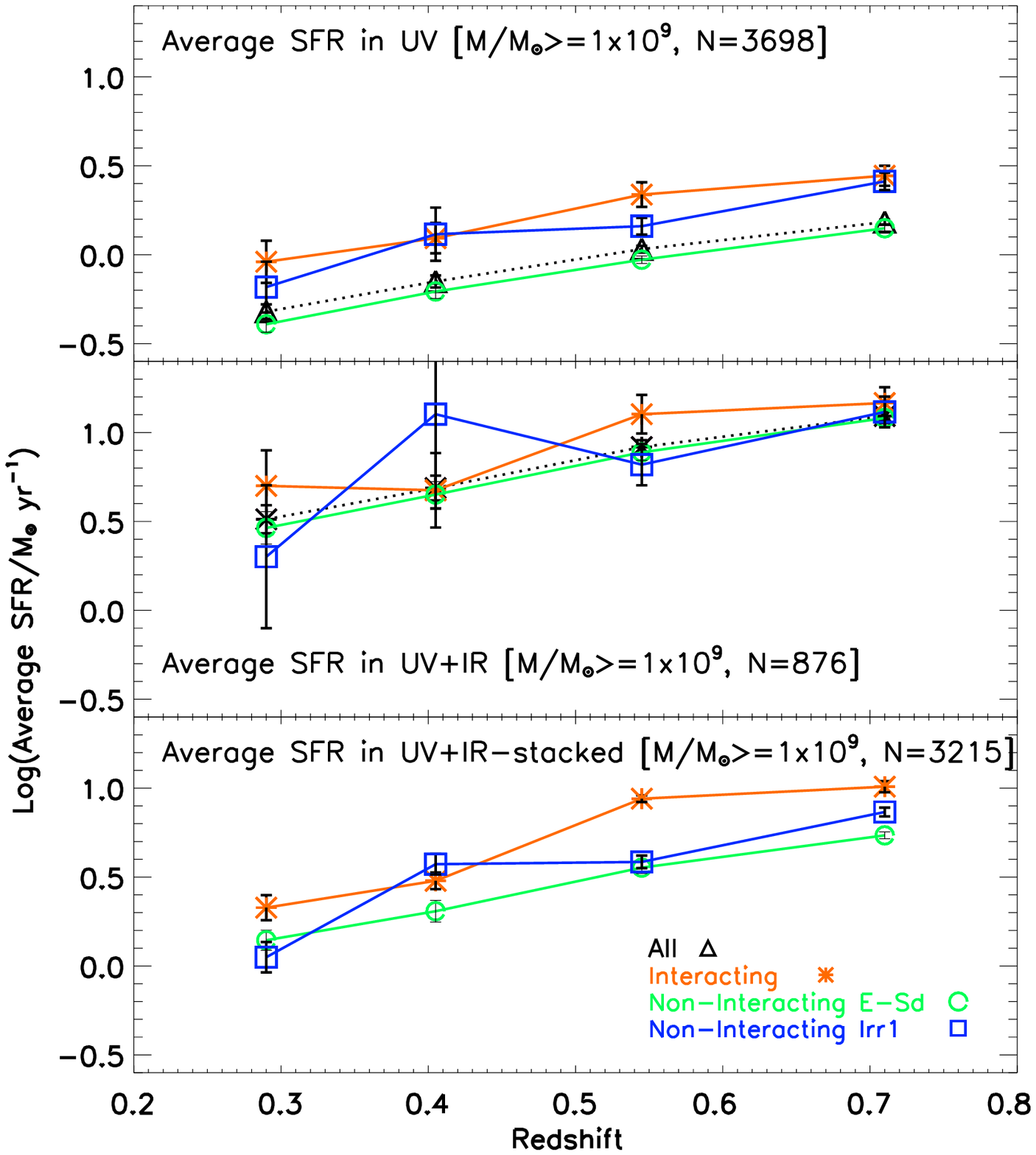}
\hspace{5.0 mm}
\includegraphics[width=2.7 in] {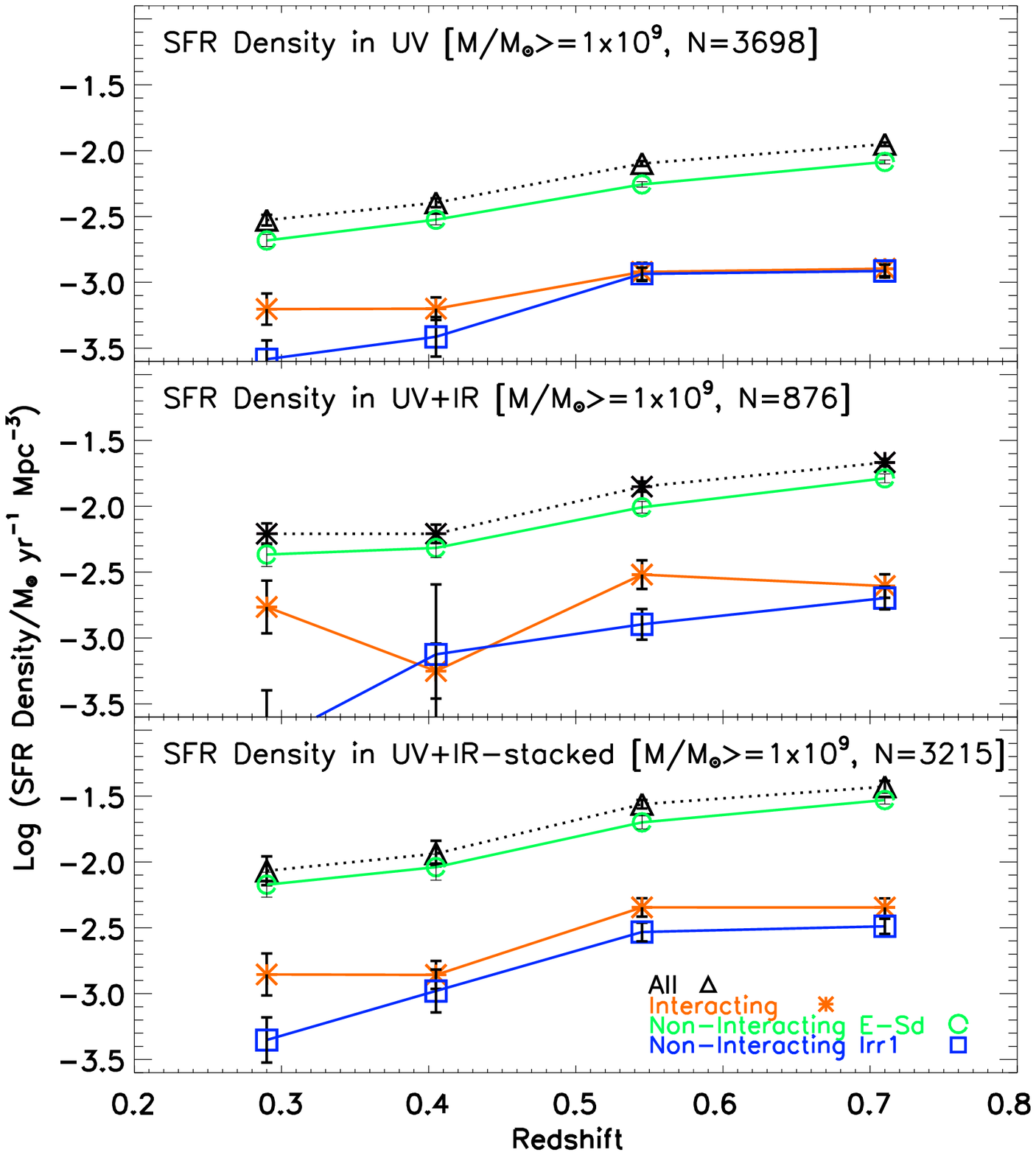}
\\
\vspace{1 mm}   
\noindent 
{\bf Fig.~1a~(Top Left):}~We show the observed  fraction of  interacting/merging galaxies  from 
Lotz et al.  (2008), Jogee  et al. (2008b), and Conselice (2003).
{\bf Fig.~1b~(Top Right):}~The empirical rate  of galaxy mergers with mass ratio $M1/M2>$~1/10 (orange stars)
among   high mass galaxies  is compared to the rate of (major+minor)  mergers (solid lines)  
predicted   by different  $\Lambda$CDM-based models of galaxy evolution. 
{\bf Fig. 1c~(Lower Left):}~The average SFR of  interacting and non-interacting  galaxies are compared.
The average UV-based SFR (top panel; based on 3698 galaxies),  average UV+IR-based SFR 
(middle panel; based on only the 876 galaxies with  24um detections), and  average 
UV+IR-stacked SFR  (based on 3215 galaxies with  24um coverage) are shown.
In all there cases, the average SFR  of  interacting galaxies is only modestly 
enhanced compared to non-interacting  E-Sd galaxies  over $z \sim$~0.24--0.80 
(lookback time $\sim$~3--7 Gyr).
{\bf Fig.~1d~(Lower Right):}~As in. 2c, but  now showing the SFR density of galaxies.
In all bins, interacting  galaxies only contribute a small 
fraction (typically below 30\%) of the total SFR density.
[All figures are from  Jogee et al (2008b)]

\clearpage
\vskip -7.0 mm      
\vskip 0.3 cm   
\vspace{0.0 mm}   
\hspace{-1.5 cm}
\includegraphics[width=2.5 in]{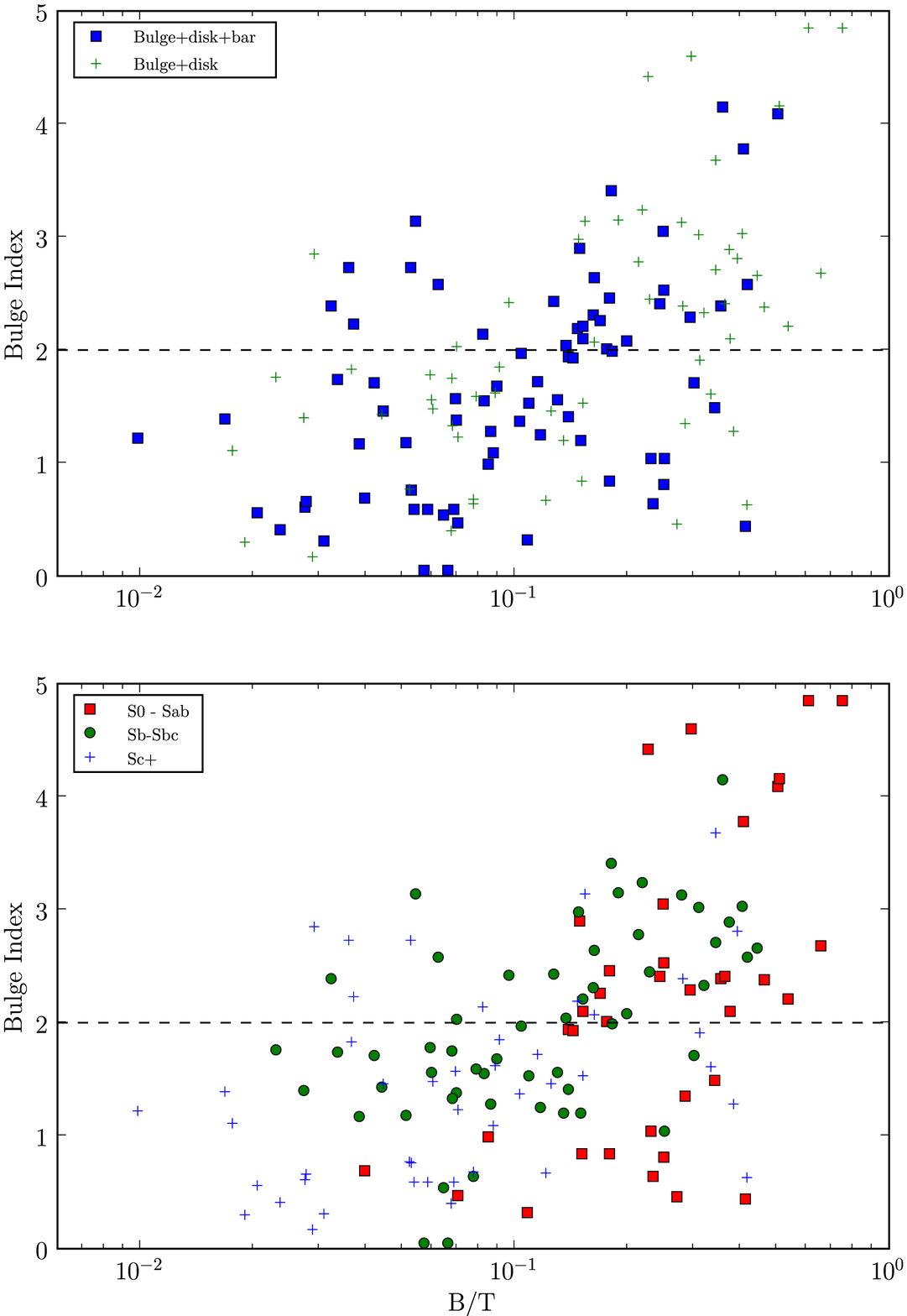}
\hspace{-1.0 mm}
\includegraphics[width=3.5 in] {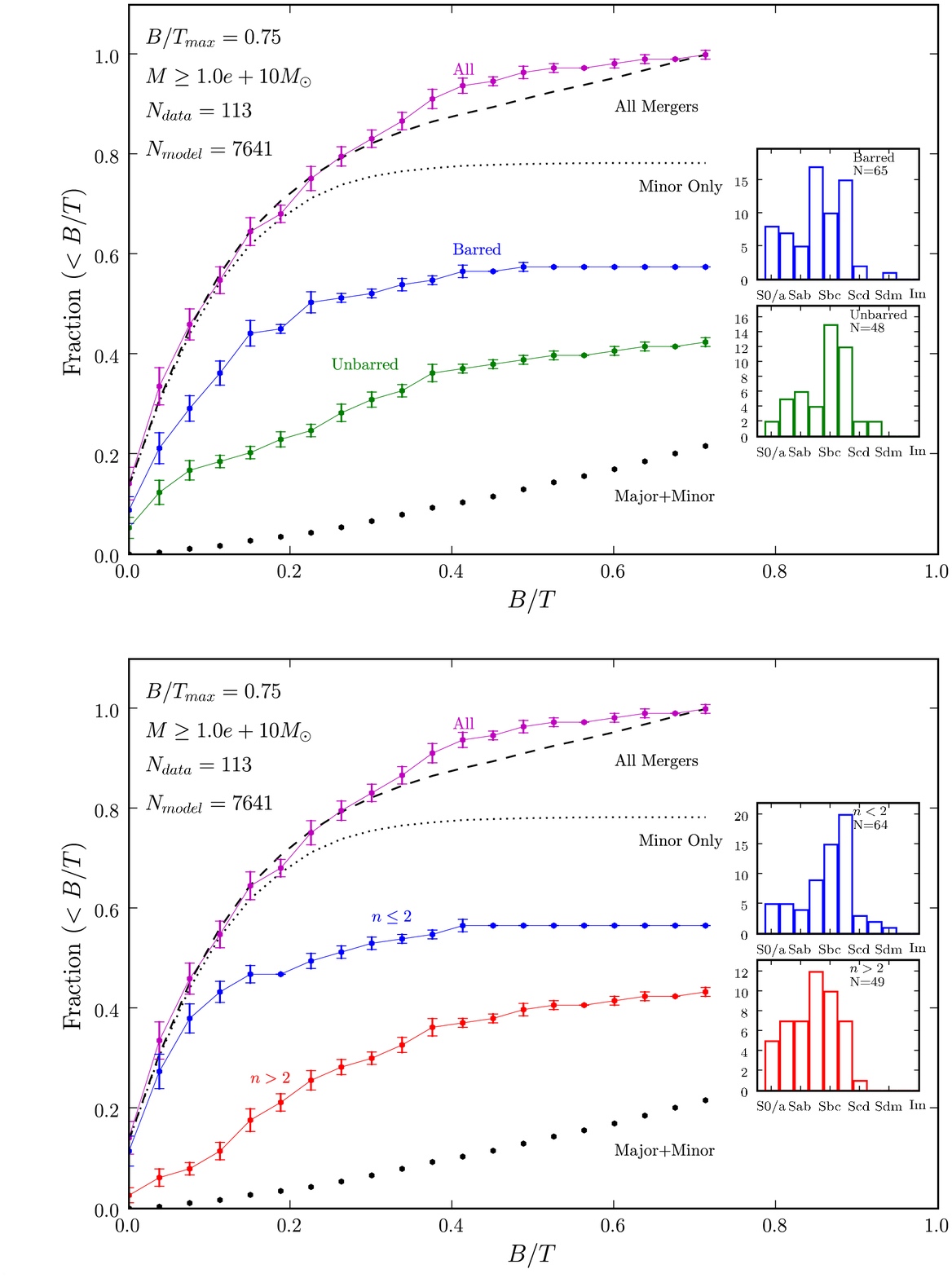}
\\
\vskip 0 mm    
\noindent 
{\bf Fig.~2a~(Left):}~The relation between $B/T$ and bulge index is shown.
The legend indicates the type of decomposition used for each data point.
Note that as many as $60\% $ of bright  spirals have 
low $n \leq 2$ bulges: such bulges exist  in barred and unbarred galaxies 
across all Hubble  types, and  their $B/T$  ranges from 0.01 to 0.4, 
with most having $B/T \leq$~0.2.
{\bf Fig.~2b~(Right):}~For high mass  ($M_\star \geq 1.0 \times 10^{10} M_\odot$) spirals, we compare the  empirical distribution of  bulge-to-total mass ratio ($B/T$) 
to predictions from  $\Lambda$CDM-based simulations of galaxy evolution. 
The y-axis shows the {\it cumulative} fraction $F$ of galaxies 
with $B/T \le$ a given value.
The magenta line shows $F$  from the data, while  the other two colored 
lines break this $F$ in terms of bar class (top panel) or  bulge $n$ (lower panel). 
The black dashed line  shows $F$  from all  model
galaxies, while  the black dotted  line and black dots show the contribution
of model galaxies that  experienced, respectively,
{\it only past minor mergers}  and {\it both major and minor mergers}.
In the models, the fraction ($\sim$~3\%) of high mass spirals, which have 
undergone a past major merger  and host a bulge  with  $B/T \le 0.2$ is  
{\it a factor of over 15 smaller}  than the observed fraction  ($\sim$~66\%) 
of high mass spirals  with   $B/T \le 0.2$.
Thus, {\it  bulges built  via major mergers seriously fail to account for most of 
the low $B/T \le$~0.2  bulges present in  $\sim$~66\% high  mass spirals}. 
[All figures are from  Weinzirl, Jogee, Khochar, Burkert, \& Kormendy (2008)]

\end{document}